\documentclass[10pt,twocolumn]{revtex4-1}
\usepackage{bm}
\usepackage{hyperref}
\usepackage{makeidx}
\usepackage{amsmath}
\usepackage{graphicx}
\usepackage{dcolumn}
\usepackage{bm}
\usepackage{color}
\usepackage{graphics}
\usepackage{amsfonts}
\usepackage{amssymb}
\usepackage{epsfig}
\usepackage{bm}
\usepackage{hyperref}
\usepackage{makeidx}
\usepackage{amsmath}
\usepackage{graphicx}
\usepackage{dcolumn}
\usepackage{bm}
\usepackage{color}
\usepackage{graphics}
\usepackage{float}
\usepackage[british,UKenglish,USenglish,english,american]{babel}
\usepackage{epstopdf}

\begin{document}

\title{Bosonic Dirac Materials on a honeycomb antiferromagnetic Ising model}

\author{\surname{Griffith} M. A. R.}
\affiliation{Departamento de Ci\^{e}ncias Naturais, Universidade Federal de S\~ao Jo\~ao Del Rei, Praça Dom Helv\'ecio 74, 36301-160, S\~ao Jo\~ao Del Rei, MG, Brazil}
\email{griffithphys@gmail.com}
\author{S. \surname{Rufo}}
\affiliation{Centro Brasileiro de Pesquisas F\'isicas, \\Rua Dr. Xavier Sigaud, 150 - Urca, 22290-180,  Rio de Janeiro, RJ, Brazil}
\author{Minos A. Neto}
\affiliation{Departamento de F\'{\i}sica, Universidade Federal do Amazonas, 3000, Japiim, 69077-000, Manaus-AM, Brazil}
\date{\today }

\begin{abstract}

Motivated by the recent proposal of Bosonic Dirac materials (BDM), we revisited the Ising model on a honeycomb lattice in the presence of the longitudinal and transverse fields. We apply linear spin-wave theory to obtain the magnon dispersion and its degenerated points. These special degenerated points emerge on the excitation spectrum as a function of the external fields and can be identified as Bosonic Dirac Points (BDP). Since that, in the vicinity of these points the Magnons becomes massless with a linear energy spectrum as well as insensible in relation to weak impurity, exactly as it occurs with a Fermionic Dirac point. We also have calculated the quantum and thermal fluctuations over the ground state of the system using Effective Field Theory. Our results point out that this simple model can host Bosonic Dirac points and therefore is a suitable prototype to build a Bosonic Dirac material only controlled by external field.

\end{abstract}

\maketitle
\section{Introduction}

The Ising model has been largely applied  in statistical physics, quantum field theory, economy and biophysics~\cite{Ising,KinrossSachdev,Tsai,Stauffer,Balatsky,Gesualdo,Alejandra,Michael, Grimmett,Bravyi,Elliott,Rudolf,Creswick,Neto1,Neto2,Kaizoji,Queiros,Burns,Owerre,Atas,Dmitriev,Takahashi1,Takahashi2,Takahashi3,Arovas,Dotsenko}. In condensed matter physics, this model can be used to study magnetic phases and its phase transitions. In particular, to study quantum phase transitions in the Ising model is necessary to add an external magnetic transverse field.
Generally, the transverse field induces a quantum phase transition from a ferromagnetic (F) or antiferromagnetic (AF) phase to a polarized one, In which the majority of the spins is along the direction of the field. In this case, the ground state before the transition (F/AF) can not be adiabatically connected with the ground state of the polarized phase.

Another intriguing result is the existence of Bosonic Dirac Materials (BDM). As pointed by Balatsky et al ~\cite{Balatsky}, the concept of fermionic Dirac materials can be extended to a bosonic version BDM. To perform this concept to a BDM is enough to show the following points; A linear characteristic of the energy dispersion in the vicinity of a point where the energy gap goes to zero; The robustness of these degenerated point against a weak disorder potential and a relative stability of these points against higher expansions of the spin wave theory.

In this work, we will demonstrate that a simple Ising model defined on the honeycomb lattice and subjected to the presence of transverse and longitudinal external field fulfils all these rules and therefore can be used as a prototype for BDM. Theoretically, the  model can be written as
\begin{equation}\label{model}
\mathcal{H}= J\sum_{<i,j>}S^{z}_iS^{z}_j -h_x\sum_{i}S_{i}^{x}-h_z\sum_{i}S_{i}^{z},
\end{equation}
such that, $J$ is the nearest neighbor exchange interaction, $S_ {i}^{z}$ is the $z$ spin component in the site $i$, $h_x$ and $h_z$ are the transverse and longitudinal magnetic fields, respectively. The Eq.~\ref{model}, was subjected to several methods and different lattices~\cite{KinrossSachdev,Grimmett, Bravyi, Elliott}.

Recent works point out that, the F and AF Heisenberg model on a honeycomb lattice exhibits degenerated points in the first Brillouin zone. At these points, the spectrum can be linearized and associated with BDP. Beside that, these BDP are robust against higher order Magnon-Magnon interaction, which are responsible to make a shift in the spectrum~\cite{Balatsky}.

The main difference between our results and the results from ref.~\cite{Balatsky} is the presence of the Bosonic Dirac points only in the polarized phase. Therefore, is possible turn on or turn off  these Bosonic Dirac Point by adjusting the external magnetic fields.

We find that by a fine tuning of the magnetic field, is possible to generate BDP located at the corners of the first Brillouin zone of the honeycomb lattice. From this, follows that is possible to transform a non-Bosonic Dirac phase into a Bosonic Dirac phase and vice verse. This is fascinating because at these points the magnons are massless, insensible to a weak disorder scattering potentials and can be propagated with very high speed $v_B$. A directly application of this result can be found in the generation of efficient spin transport~\cite{Pires}. It is worth to note, that some mechanism of the creation, localization and vanishing of these Dirac-like points, has attracted theoretical and experimental researchers~\cite{Rechtsman,Lu1,Lu2,Wang,Wang2,Kahanikaev,Dietl,Montambaux,Tarruell,Bellec}.

In addition, we determined the effects of the competition between the transverse and longitudinal fields, the influence of
on-site potential disorder over the BDP, edge states behavior and temperature effects.

For this purpose, we used two different methods, linear spin-wave theory (LSWT) and effective-field theory 1 (EFT-1).

The LSWT has been used to build the ground state and the dispersion in the real and momentum space.
In the real space, with open boundary conditions promoting a zig zag termination, we showed the edge states for a clean and disorder affected systems.
On the other side, EFT-1 has been used to compute zero and finite temperature effects over the sublattice magnetization, as a generalization of the pervious works~\cite{Creswick,Neto1,Neto2}.

Experimentally, in order to test our results we suggest the scanning tunneling microscopy (STM) technique. Firstly, STM can be used to create a $2D$ honeycomb magnetic lattice~\cite{Gomes} by depositing atoms or molecules in substrates. Secondly, the STM is able to change continuously the magnetic exchange interaction simply manipulating the position of the absorbantes (atoms, molecules etc..) which are deposited in a subtraction. Such that, for the present purpose an external field can be added to verify the controlling over the BDP.

The paper is organized as follows; In the section~\ref{PhaseDiagram}, we will discuss about the relevant ground states; The excitation spectrum  critical exponents, Bosonic Dirac Point will be discussed in the section~\ref{LinearBehaviour}, while the impurity effects and edge states has been calculated in the sections \ref{Impurities} and \ref{Realspace}, respectively. The magnetic properties and finite temperature effects will be showed in section~\ref{FiniteTemperature}. Finally, in the last section~\ref{conclusion}, we summarize our results.

\section{Phase Diagram and ground state}
\label{PhaseDiagram}

\begin{figure}[h!]
\centering
\includegraphics[scale=0.8]{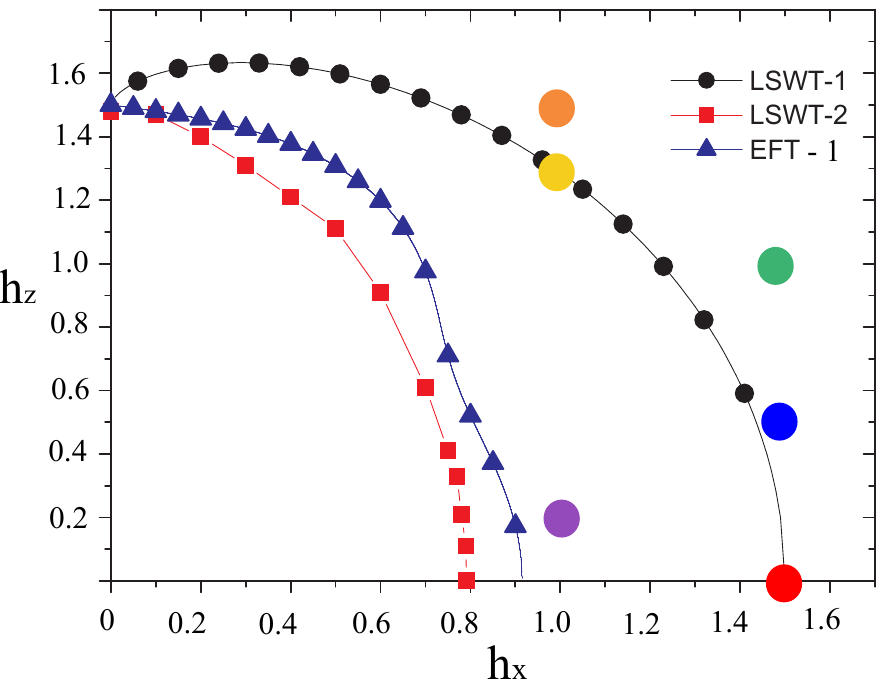}
\caption{(color online) Phase diagram at zero temperature in plane $(h_x,h_z)$. Each critical line represents a phase transition from the canted antiferromagnetic to a polarized phase. Black circles, red squares and blue triangles were obtained by LSWT-1, LSWT-2 and EFT-1, respectively. The purple, yellow, orange, red, blue and green big solid circles has been used to calculate the magnon energy dispersion in the Fig.~\ref{Dispersionplane}.}
\label{phasediagram}
\end{figure}

\begin{figure}[ht!]
\centering
\includegraphics[scale=0.9]{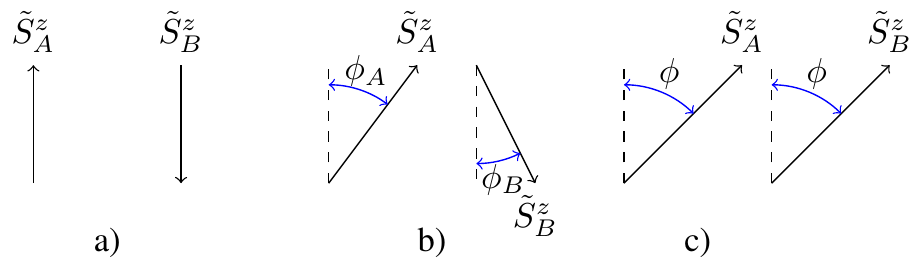}
\caption{(color online) Spin configurations. a) Antiferromagnetic states in absence of the magnetic fields, where $\phi_A=\phi_B=0$. b) Canted antiferromagnetic phase, where $\phi_A \neq \phi_B$. c) Polarized phase, where $\phi_A=\phi$ and $\phi_B=\pi-\phi$ or $\phi_A+\phi_B=\pi$.}
\label{angles}
\end{figure}

Using LSWT and EFT-1 (appendix \ref{spinwave} and \ref{EfectiveFieldTheory}, respectively), we determined the phase diagram of the system in plane $h_x-h_z$, see Fig.~\ref{phasediagram}. The critical lines separates a canted antiferromagnetic (CAF) (region below the critical line) phase from polarized phases (PP)
(region above the critical line), see Fig.~\ref{angles}. The black and red lines were obtained by LSWT, while the blue line is a result from EFT-1.

We obtained the black line minimizing the classical energy $E_g=E_c$ with respect to angles $\phi_A$ and $\phi_B$ (LSWT-1). On the other side, adding the quantum correction $E_q$ in the the energy, we get the red line (LSWT-2). Note that, at the critical line and in the region above the critical line $\phi_A+\phi_B  =  \pi$ (Fig.~\ref{angles} (c)).

Still about the angle spin configuration, see the quantum phase transition that occurs at $h_{x} =1. 5$ for $h_{z}=0$ (red big solid circle in Fig.~\ref{phasediagram}). Above $h_{x} \geq 1.5$ the spins are totally along the x-axis. Following, for $h_x=1.5$ and increasing the longitudinal field from zero to $h_{z}=0.5$ (red big solid circle) the spin configuration of the polarized phase is no more totally along the x-axis, getting an effective angle $\phi$ as in Fig.~\ref{angles} (c). Along the critical line, this angle changes and finally for $h_{x}=0$ and $h_{z}=1.5$, $\phi=0$.

The results from LSWT-2 are qualitatively equal to LSWT-1, in the sense that it describes a continuous transition between CAF and P phases. However, one can observe a considerable reduction of the CAF region for LSWT-2 in comparison to LSWT-1, see Fig~\ref{phasediagram} (red line and triangle). This is a new result, in particular in the case of a Honeycomb lattice and corroborates with the idea that the spin wave theory tends to overestimate the ordered phase region in a first approach.


The critical line from EFT-1 (blue line and triangles) was also included Fig.\ref{phasediagram} and seems to be in great agreement with the critical line obtained from LSWT-2 (red squares). Here, the EFT-1 method takes into account a cluster of one spin interacting with their vicinity. We considered a staggered magnetization as a order parameter and the quantum critical point can be founded throughout the Eqs.~\ref{B15}-~\ref{B16}. It is worth to communicate that LSWT-1 provides results quantitative equal to a first approximation of Bogoliobov Mean Field Theory (MFA-1) for a cluster with one spin \cite{Neto2}.

\begin{figure}[t]
\centering
\includegraphics[width =3.5in]{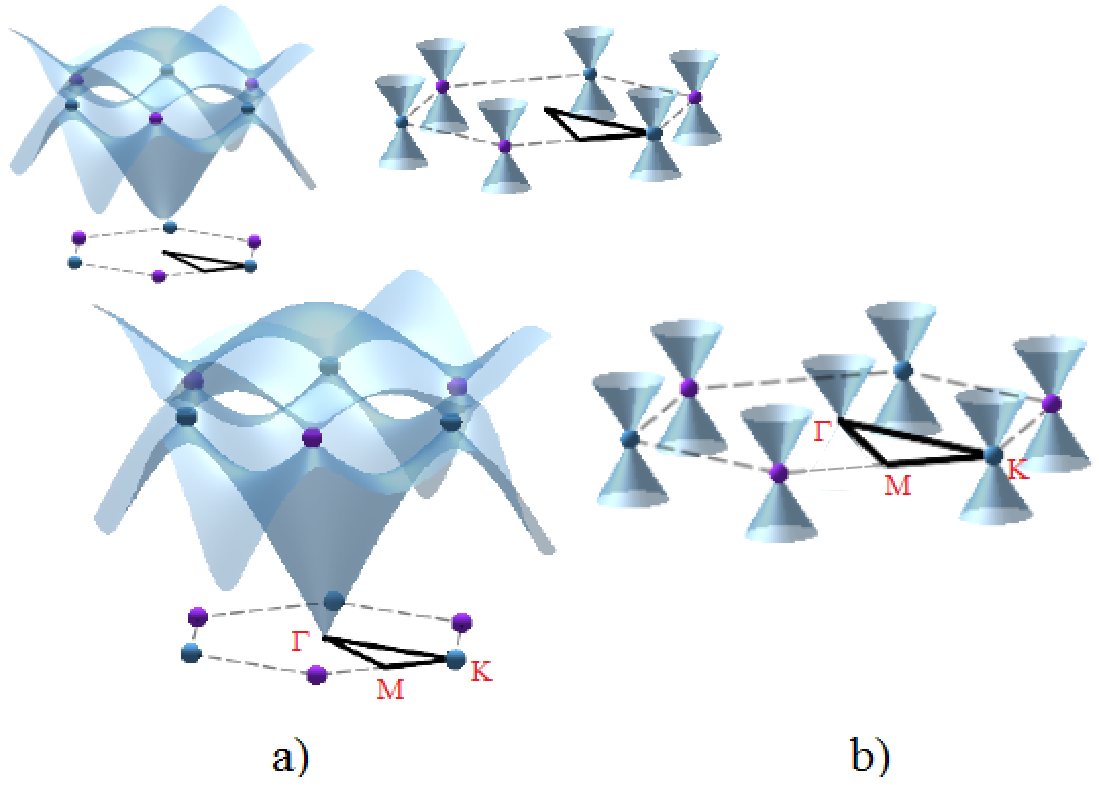}
\caption{(color online) Quasiparticle excitations and Bosonic Dirac cones.
a) Bosonic Dirac points within the polarized phase $h_x=1.5,h_z=0.5$ (top) and at the critical point $h_x=1.5,h_z=0.0$ (bottom).
b) First BZ and the locations of the Bosonic Dirac cones at the corners $K$ (bottom).
Here  $\Gamma=(0,0)$ (soft mode) , $M=( \frac{2\pi}{3a},0)$ and $K=( \frac{2 \pi}{3a}, \frac{2\pi}{3\sqrt{3}a})$ are points in the BZ for a lattice parameter $a$, which define the path shown in the BZ.}
\label{DispersionBZ}
\end{figure}

\section{Linear Behaviour, Bosonic Dirac points and critical exponents}
\label{LinearBehaviour}

The linear behaviour can be identified around the special points called by $\Gamma=(0,0)$ and $K=( \frac{2 \pi}{3a}, \frac{2\pi}{3\sqrt{3}a})$ within the first Brillouin zone, see Fig.~\ref{DispersionBZ}. Only the $K$ points are possible candidate to be  ``Bosonic Dirac points", following the same nomenclature adopted by the Balatsky et al. \cite{Balatsky}.

The low energy description of the magnon spectrum provides a linear dispersion around $\Gamma$ equal to $\omega(\mathbf{\Gamma}) \sim  D \sqrt{ \Delta^{2\nu z } +  k^{2z} }$. Where the gap function $\Delta$ goes to zero at the critical point as well as $\nu$ and $z$ are the correlation and dynamical critical exponents. Note that, $D$ and $\Delta$ are independent of $k$, they are given by $D=\sqrt{ \frac{{\tilde{A} J z S {\sin(\phi_A)}^2}}{4}}$ and $\Delta=\frac{\sin(\phi_A)^2}{D^2} \left( h_{x} - h_{xc} \right)$ (since we choose to fix $h_{z}$),  respectively.
The critical value can be written as $h_{xc}=\frac{JzS-h_{z}\cos(\phi_{A})}{\sin(\phi_{A})}$ where the gap vanishes, as shown in more details in appendix~\ref{Lowenergy}. These calculations leading to $\nu=\frac{1}{2}$ and $z=1$ for the critical exponents, the last ensure the linear behaviour.

For the critical exponent of the correlation length, we found $\nu=\frac{1}{2}$ as expected for a mean field approach. It is worth mentioning, for a 2D Ising model with a transverse field the value gets $\nu=0.63$~\cite{Elliott,Stinchcombe}.

Following, around the $K$ the dispersion gets a linear behavior as $\omega(\mathbf{K}) \sim \tilde{A} \pm  \frac{3}{4} J S \sin(\phi_A)^2 \delta k$. It is worth to note that in this case the gap function is $ \Delta = (\omega_1(\mathbf{K})-\omega_2(\mathbf{K}))$, in particular, for $h_z=0$, $\Delta= (h_{xc}- h_x )^{\nu z}$~\cite{Stinchcombe,Croo} with $\nu z =1$.

The $z=1$ result for the Ising model in a presence of transverse, is known and works for any regular lattice since the dimension of the system be an integer~\cite{SSachdev}. In addition, this is also independent of the quantum critical point as demonstrated for the $\Gamma$ and $K$ points.

\begin{figure}[t]
\centering
\includegraphics[scale=0.46]{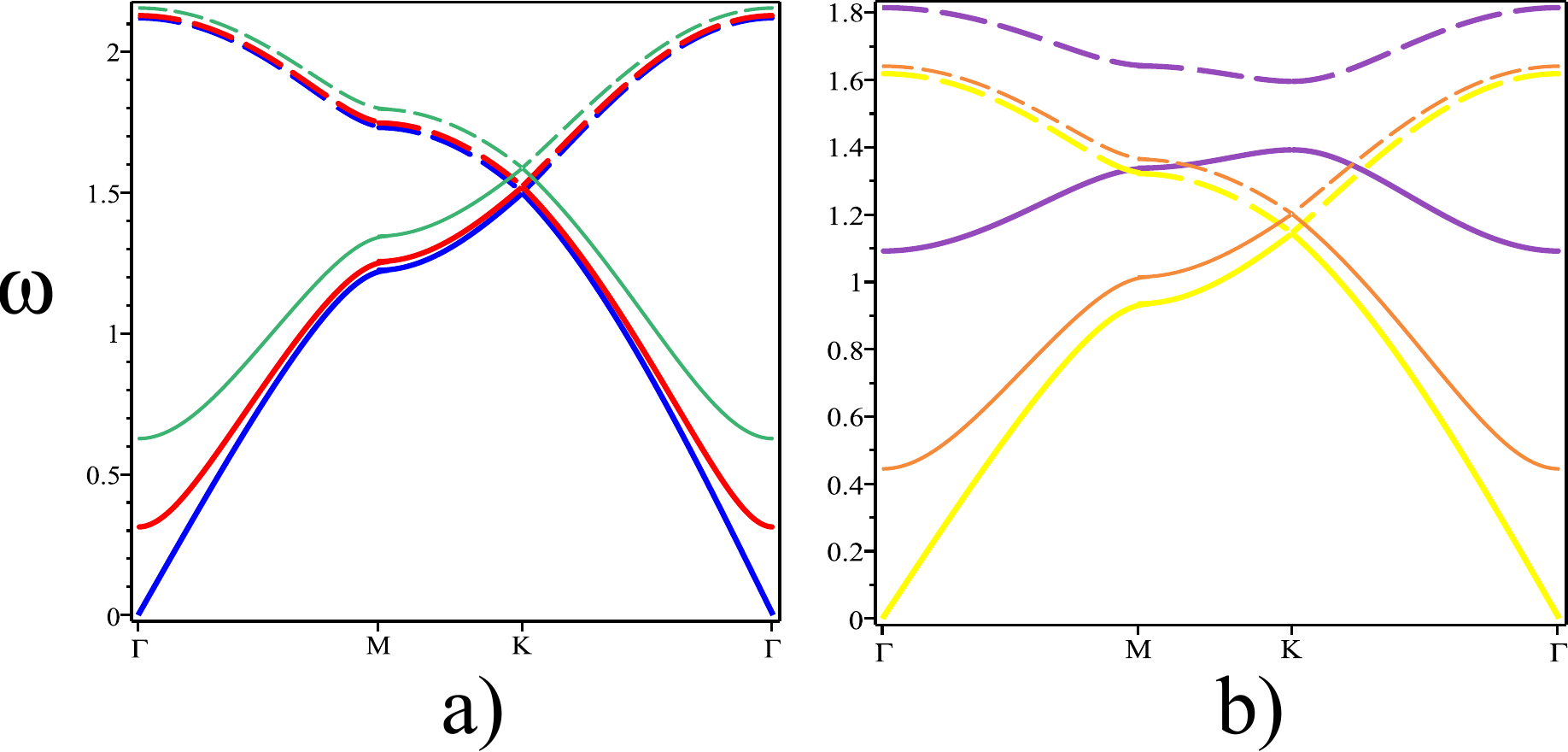}
\caption{(color online) Magnon quasi-particle excitations  $\omega_{1}$ (solid) and $\omega_{2}$ (dashed) along the path $\Gamma-M-K-\Gamma$.
a) We have fixed the transverse field at $h_x=1.5$.  Curves for longitudinal field $h_z=0.0$ (red), $h_z=0.5$ (blue) and $h_z=1.0$ (green).
In the case of red curve, the system is gapless and we can see a linear behaviour that emerges at the vicinity of the $\Gamma$ and $K$ points.
Otherwise, in case of blue and green curves, the linearity in dispersion only occurs near the $K$ point.
b) Now, the transverse field has been fixed at $h_x=1.0$ and we have calculated the magnon dispersion for $h_z=0.2$ (purple), $h_z=1.2871$ (yellow) and $h_z=1.5$ (orange). }
\label{Dispersionplane}
\end{figure}

We are reporting in this work that theses linear behaviours appear in spectrum as a function of the external fields. The dominant set of critical exponents is dictated by the soft point at $\Gamma$.
The number and location of these special points depends on criticality and the magnetic phase, see Fig.~\ref{Dispersionplane} and Fig.~\ref{phasediagram}.

These results can be summarized as;
In the CAF phase there is no Bosonic Dirac Points, sice the system is gapped. At the critical line, the energy dispersion becomes gapless and six Bosonic Dirac points emerge at corners of the Brillouin zone ($K$) and there is a linear soft mode at the center ($\Gamma$). These six Bosonic Dirac point at the corners of BZ persist in the polarized phase.

Our report is quite different from the case discussed in ref.~\cite{Balatsky}, since the existence of the Bosonic Dirac Points depends on the magnetic phase. For the present case, the controlling of these Dirac points resides in a simple adjusting of the magnetic fields instead of the exchange interaction $J$.

In Fig.~\ref{Dispersionplane}, we explore the projection of the magnon excitation along of the path defined by $\Gamma-M-K-\Gamma$. This figure shows the spots of the Bosonic Dirac points in the BZ and how they are affected by the magnetic fields.

The color scheme follows the big solid circles in Fig.~\ref{phasediagram} for each pair $h_{x}-h_{z}$. The red line in the Fig.~\ref{Dispersionplane} a), is the quasiparticle excitations at the critical point $(h_x=1.5,h_z=0)$. For a fixed $h_x=1.5$ and varying $h_z$ from zero to 0.5, we get the point $(h_x=1.5,h_z=0.5)$ (polarized phase), where the dispersion is linear only around the $K$ points. The same behavior occur at the point $(h_x=1.5,h_z=1.0)$ (green curve). A similar behavior can be observed, but for a fixed $h_x=1.0$, see Fig.~\ref{Dispersionplane} b). Therefore, the Bosonic Dirac points seems to be a particular characteristic of the polarized phase.

\section{Impurity effects and impurity resonance}
\label{Impurities}

Are these degenerate Bosonic Dirac points robust against impurities? To answer this question, we investigated the effects of an impurity scattering. The impurity simulates a local defect created by any removing/modification in the spin over the honeycomb lattice. For this reason, we considered a defect given by $H_{imp}=\sum_{i} V_0 a^{\dagger}_i a_i $.

In order to present the physical conditions that the impurity scattering should obey to generate an impurity resonance, exactly at the Bosonic Dirac point $K$ and linear point $\Gamma$,  we need to calculate the Green's function of the system from the Dyson equation. The absence of an impurity resonance at these points implies robustness against the scattering potential $V_0$. So, this robustness need to be checked in order to make the association with ``Bosonic Dirac point".
We write the impurity Green's function using the T-matrix approach. The single magnon Green's function for our problem is given by
\begin{equation}\label{SingleGreensfunction}
  G(k,\omega) = g_0(k,\omega) +  g_0(k,\omega) T_{imp} g_0(k,\omega),
\end{equation}
where, $g_0=( \omega+ i\delta - \mathcal{H} )^{-1}$, $T_{imp}=\frac{1}{2 (V_0^{-1} + \bar{g}_0) } \left( \sigma_{0} \otimes \sigma_{0} +  \sigma_{0} \otimes \sigma_{z} \right)
$, $\sigma_i $ are the usual Pauli matrix and $\bar{g}_0 (\sigma_{0} \otimes \sigma_{0})=\frac{1}{N}\sum_k  g_0(k,\omega)$.

\begin{figure}[t]
\centering
\includegraphics[scale=0.45]{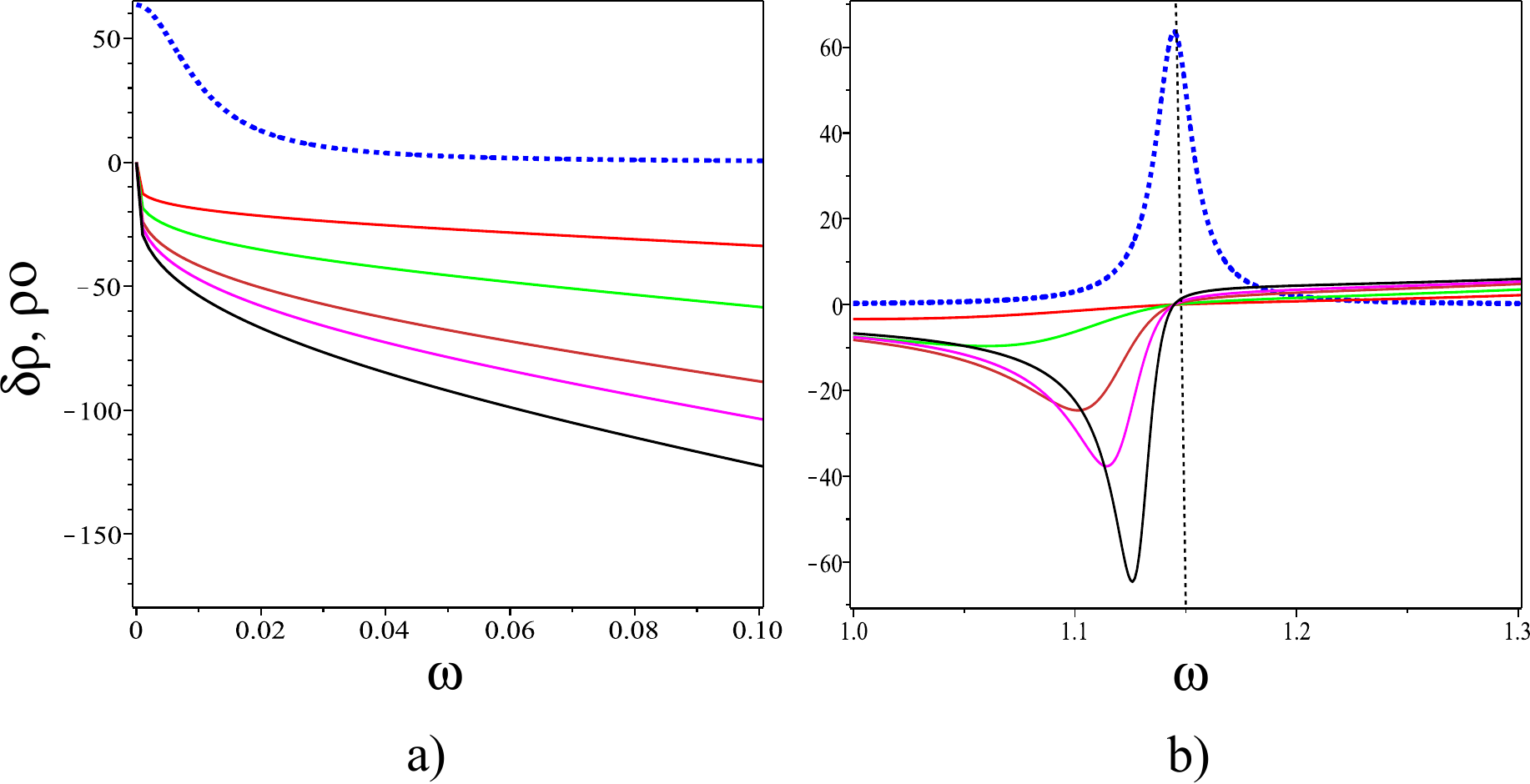}
\caption{(color online) Density of state $\rho_0(\omega)$ and impurity correction $\delta \rho(\omega)$ at critical point A $(h_x=1.0,h_z=1.28)$. Dashed blue and solid lines represents $\rho_0(\omega)$ and $\delta \rho(\omega)$, respectively. The calculations were performed
a) around $\Gamma$ point and  b) around K point. In both cases, the scattering potential assumes $V_0=10,20,50,10^2,10^4$ following the solid lines from top to bottom order according with the curves (red, green, dark red, mangueta and black curves).}
\label{Densityofstates}
\end{figure}

To determine the impurity effects over the phase diagram in Fig.~\ref{phasediagram}, we chose two points A and B; The A point is given by $(h_x=1.0,h_z=1.28)$, which corresponds to a critical point. The B point $(h_x=1.5,h_z=1.0)$ within the polarized region. Using the A point, we calculated the density of states for a clean case $\rho_0(\omega)=-\frac{1}{\pi} Im(g_0(\omega))$ (dashed blue lines of Fig.~\ref{Densityofstates}) and the impurity correction $\delta \rho(\omega) = -\frac{1}{\pi} Im(g_0(\omega) T_{imp} g_0(\omega) ) $( solid lines of Fig.~\ref{Densityofstates}). So, in the Fig.~\ref{Densityofstates} an impurity resonance might emerges only in the case of very strong values of scattering potential $V_0$. This means that the correction in $\rho_0$ is precisely zero for low values of potential scattering $V_0$. Note yet, that the pick of $\delta \rho $ approaches from the resonance limit only when $V_0$ goes to infinity. This behaviour was observed along all critical line, ergo the critical line is robusts against this kind of defect.

\begin{figure}[t]
\centering
\includegraphics[scale=0.5]{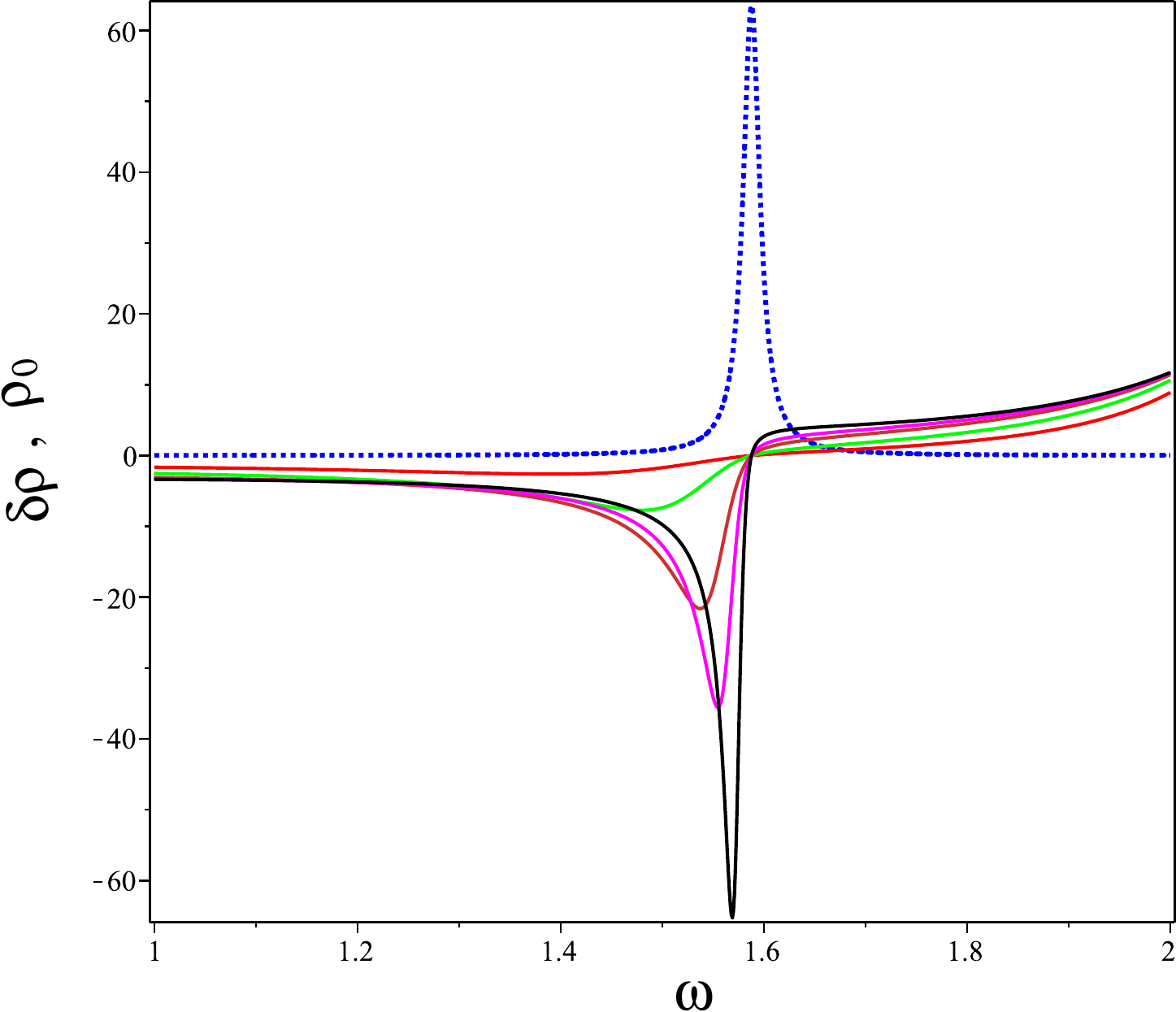}
\caption{(color online) Density of state $\rho_0(\omega)$ and impurity correction $\delta \rho(\omega)$ in the polarized phase, B point $(h_x=1.5,h_z=1.0)$ around the K. The solid lines were obtained for $V_0=10,20,50,10^2,10^4$ and follow the same color scheme of Fig.~\ref{Densityofstates}.}
\label{DensityofstatesPolarized}
\end{figure}

Furthermore, the Fig.~\ref{DensityofstatesPolarized} shows the effects of an impurity scattering over de Bosonic Dirac point $K$ in the polarized phase. Again, low values of the impurity potential $V_0$ can not generate a resonance impurity. These results can be used to guarantee that these special degenerated points $\Gamma$ and $K$ are robust against a weak impurity  potential in a similar way what as it occurs with their cousin, Fermionic Dirac points.

In the strong limit of scattering potential, the impurity resonance emerges only in the sublattice that does not hosts the scattering center. As discussed in ref.~\cite{Balatsky}, the scattering induces Friedel oscillations and the local magnon density $\rho(r,\omega)$ around the impurity describes waves emanating from the scattering center with the standard asymptotic decay proportional to $r^{-1}$.

\section{Real space and edge states}
\label{Realspace}

\begin{figure}[t]
\centering
\includegraphics[scale=0.5]{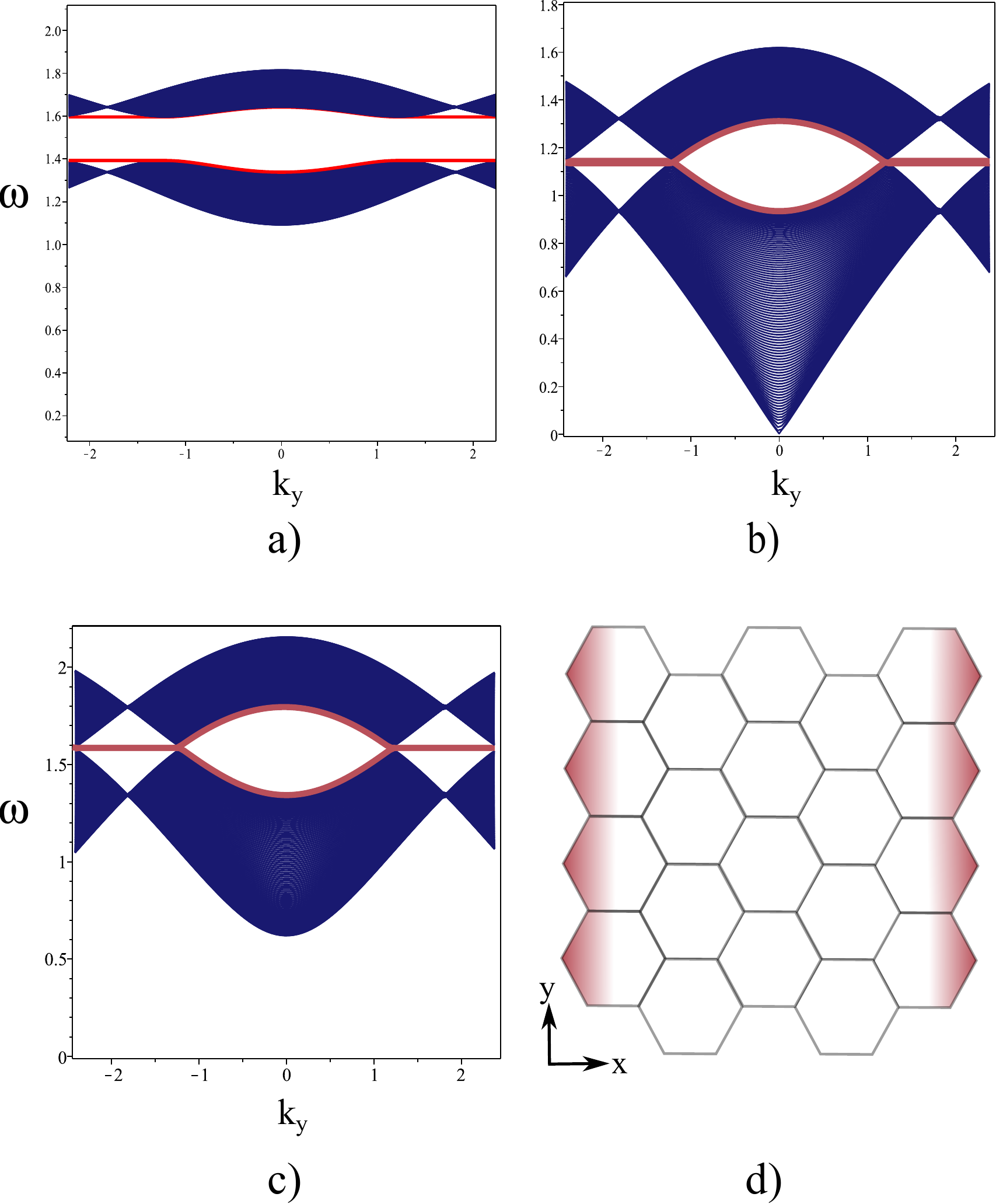}
\caption{(color online) Eigenvalues and edge states for $L_x=100$ as a function of $k_y$. The energy of the edge state is represented by red lines, while the blue lines the bosonic bulk states. a) In the CAF phase at point $h_x=1.0$ and $h_z=0.2$,  b) at the critical point $h_x=1.0$ e $h_z=1.28$ and c) in the polarized phase at $h_x=1.5$ e $h_z=1.0$. d) A pictorial representation of the zigzag termination and the location of the edges states (red gradient).}
\label{RealCleanfig}
\end{figure}

An interesting feature that arises in the finite honeycomb lattice is the presence of edges states. In this section we are devoted to understand the nature of these edges states and how a disorder potential can destroys it. The presence of the Bosonic Dirac Points at $K$ points in this finite lattice is also a relevant point to be observed.

We have written the  Hamiltonian eq~\ref{modelSpinwave} in the real space with open boundary conditions along the $x$ direction, while in the $y$ direction we considered a periodic boundary condition (see appendix \ref{HRealSpace}). Thus, the lattice becomes a kind of a finite cylinder with zigzag terminations, in a very similar way to a nanoribbon in graphene context. The edge states are localized around the zigzag terminations of this cylinder, the highlighted red region, as depicted in Fig.~\ref{RealCleanfig} d).

Without disorder, the eigenvalues of the clean Hamiltonian in real space Eq.~\ref{HbarkyZZ} are shown in the Fig.~\ref{RealCleanfig} a), b) and c) for $L_{x}=100$. Where $L_{x}$ counts the number of unit cell, containing a pair of sublattice A and B, along the finite $x$ direction.
In Fig.~\ref{RealCleanfig}, we have the bulk bosonic states (blue lines) and the zigzag edge states (red lines) of all magnetic phases.
The edges states from CAF and polarized phase are shown in the parts a) and c), respectively. In the part b), we showed how these edges states behaves at the critical line.
The location of the edge states are depicted in the part d). They are majority distributed around the terminations of the lattice as indicated by the red gradient color.

\begin{figure}[t]
\centering
\includegraphics[scale=0.5]{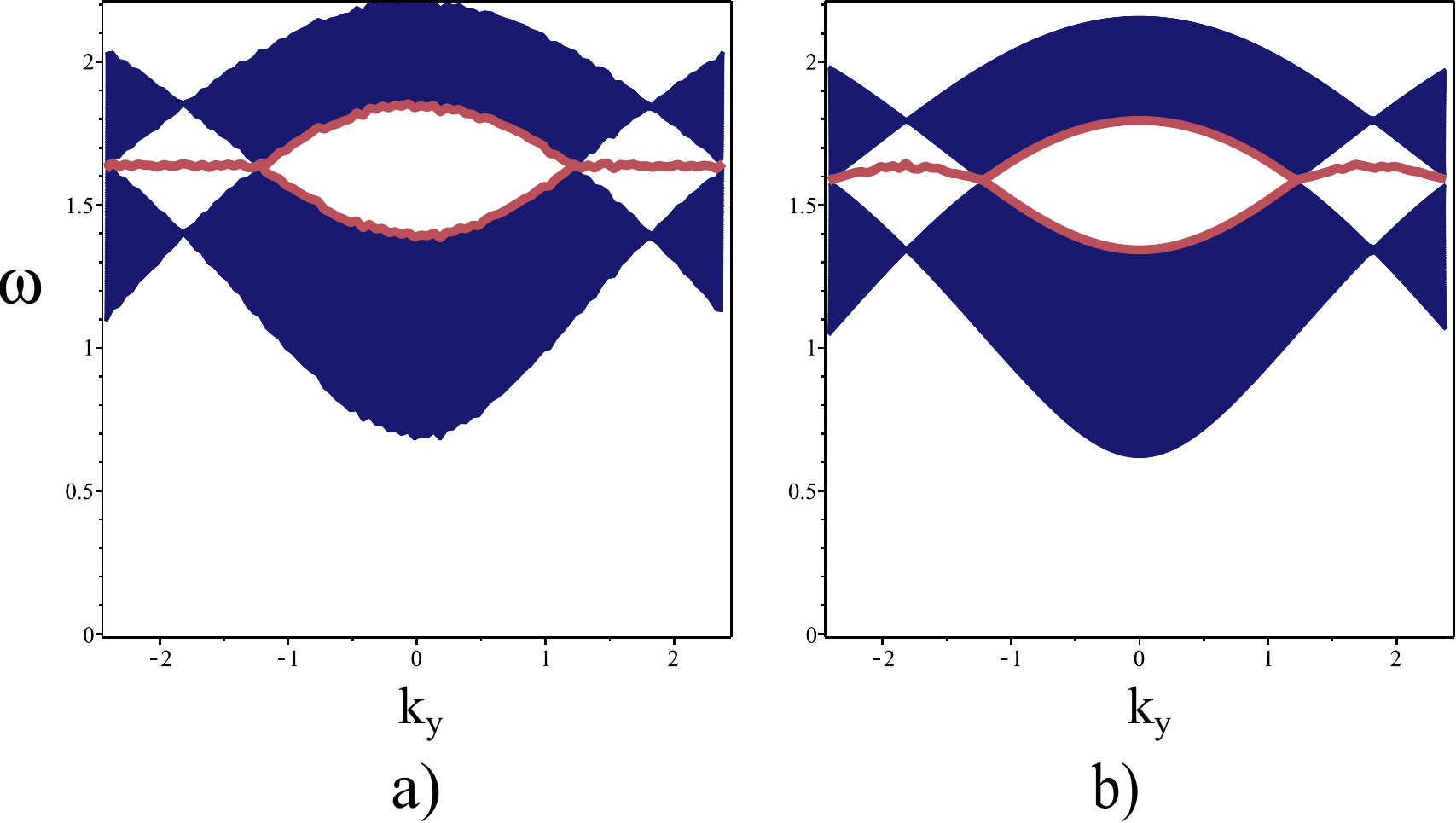}
\caption{(color online) Eigenvalues and edge states for $L_x=100$ as a function of $k_y$ for a potential disorder ensemble with 30 realisations. Calculations performed in the polarized phase at point $h_x=1.5$ and $h_z=1.0$. The energy of the edge state are represented by red lines, while the blue lines are bosonic bulk states.
a) Disorder at bulk and edges.  b) Disorder only at the edges (zigzag) of the honeycomb lattice.}
\label{RealCleanfigDisorder}
\end{figure}

A coupling with a disorder potential can simulates an interaction with a substrate. The adding of this disorder potential allow us to investigate
its effect over these Bosonic Dirac points and edge states within the polarized phase. Also, providing an another way to determine the robustness of the BDP against disorder inside a finite lattice study.

After this disorder calculations, we present the result of Fig.~\ref{RealCleanfigDisorder} for an ensemble with 30 realisations.
As we can see, a weak on-site potential disorder for both bulk and edges, and only concentrated at the edges, respectively Fig.~\ref{RealCleanfigDisorder} a) and b), can not open a gap around the points K. So, even in this finite lattice study, these $K$ points are insensible in relation to this kind of disorder and so far they prove to be robust.

\section{The temperature effects}
\label{FiniteTemperature}

Beyond the zero temperature results and in order to design a more complete study, we also investigate the behavior of the ground state in the presence of thermal fluctuations. Through the sublattice magnetization $m^z_ {A, B} $ as a function of the reduced temperature $T=k_ {B} T/J$ and the magnetic fields we developed nonzero phase diagrams.

\begin{figure}[t]
\centering
\includegraphics[scale=0.8]{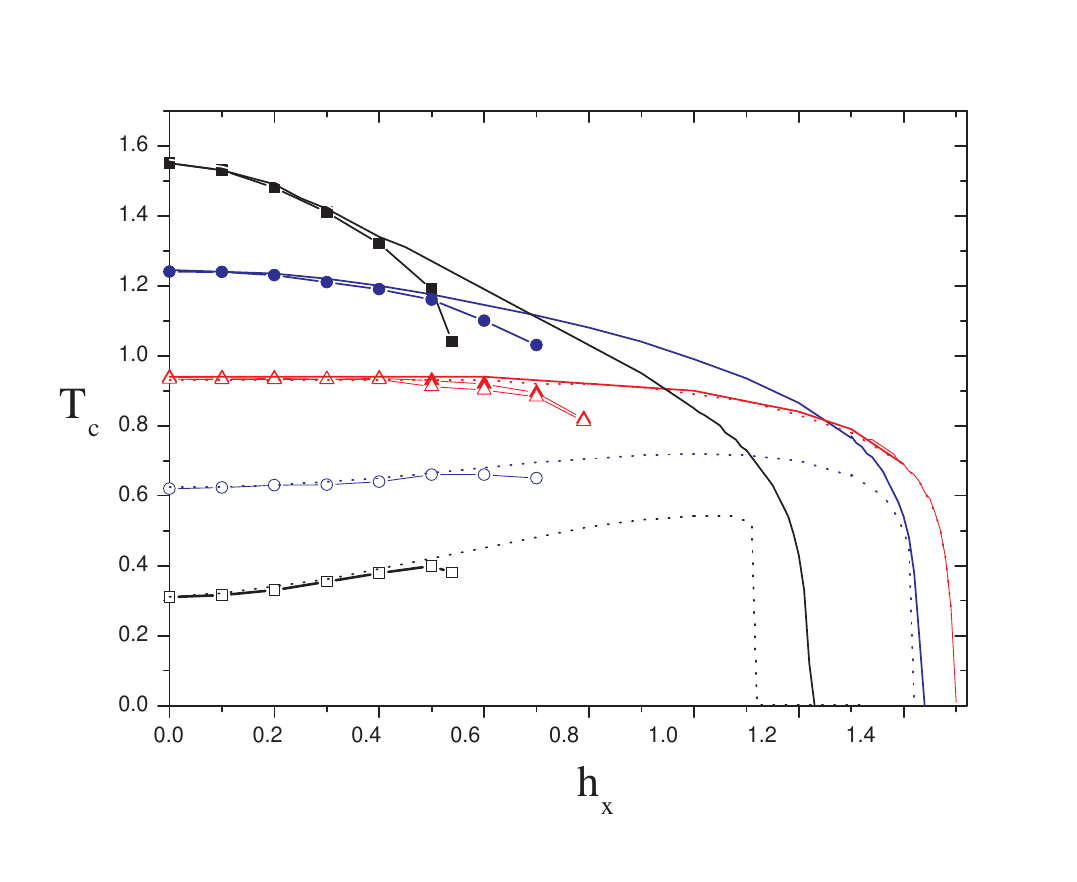}
\caption{(color online) Critical temperature as a function of $h_x$ for several values of the $h_z$ obtained by LSWT-1,2.
The LSWT-1 results for sublattice $A$ and $B$ are represented by solid and doted lines, respectively. The LSWT-2 results are signed by curves with geometrical forms solid for $A$ and opened for $B$ sublattices. The longitudinal field assumes the fixed values, $h_z=0.1$ (red), $h_z=0.5$ (blue) and $h_z=1$ (black).}
\label{TCvSHXEPS}
\end{figure}

\begin{figure}[t!]
\centering
\includegraphics[scale=0.8]{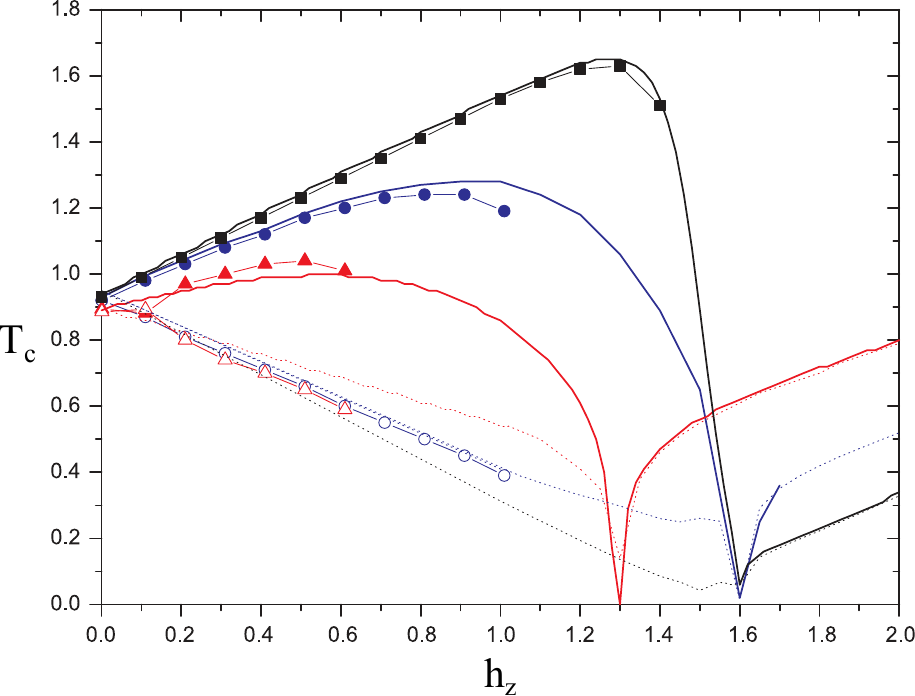}
\caption{(color online) The same of Fig.~\ref{TCvSHXEPS}, but varying $h_z$ for $h_{x}=0.1$ $h_{x}=0.5$ and $h_{x}=1$ fixed values.}
\label{TCvSHZEPS}
\end{figure}

\begin{figure}[t]
\centering
\includegraphics[scale=0.8]{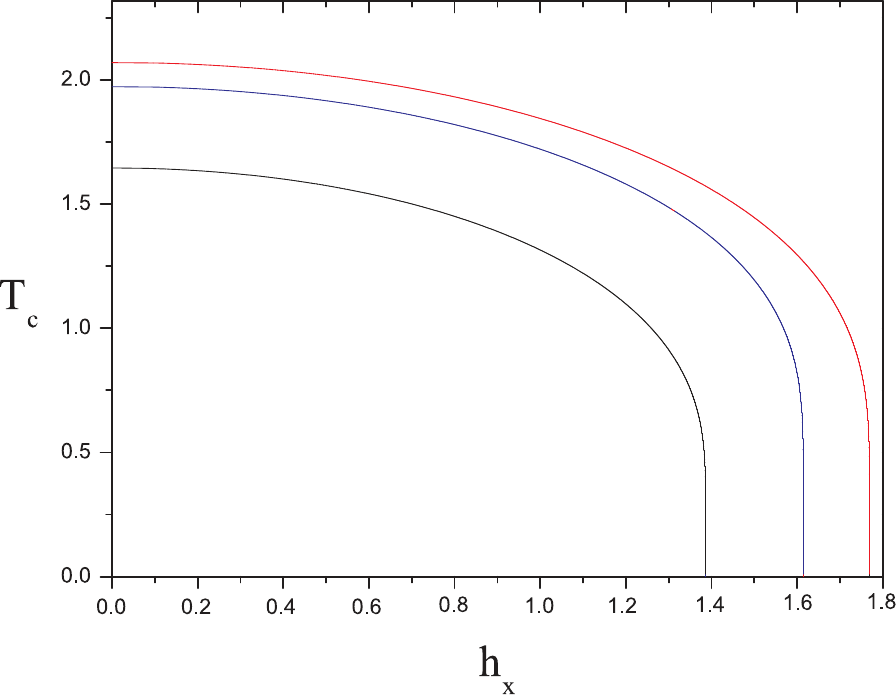}
\caption{(color online) Critical Temperature as a function of the $h_x$, obtained by EFT. For $h_z=0.5$ (red), $h_z=1$ (blue) and $h_z=2$ (black) fixed values. To compare this results with Fig.~\ref{TCvSHXEPS} the values of fields must be renormalized by $1/2$.}
\label{TcVHxEFTEPS}
\end{figure}

\begin{figure}[t]
\centering
\includegraphics[scale=0.8]{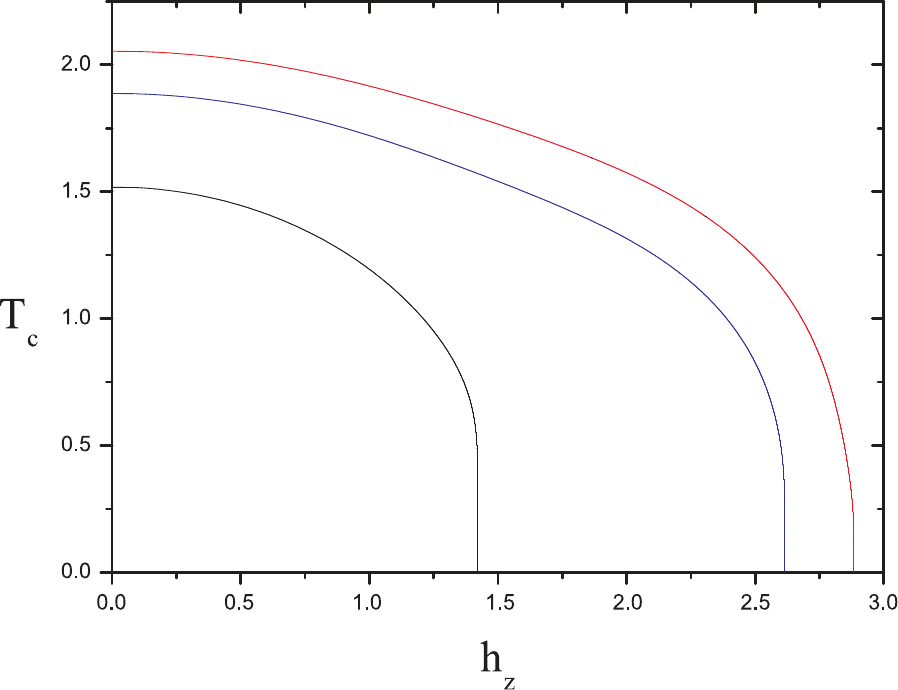}
\caption{(color online) The same of Fig.~\ref{TcVHxEFTEPS}, but varying $h_z$. For $h_x=0.5$ (red), $h_x=1$ (blue) and $h_x=1.5$ (black) fixed values. To compare this results with Fig.~\ref{TCvSHZEPS} the values of fields must be renormalized by $1/2$.}
\label{TcVHzEFTEPS}
\end{figure}

Taking into account only the critical behavior, we present the critical reduced temperature as a function of the $h_ {x}$ for several values of $h_ {z}$ obtained by LSWT, and vise versa in Fig.~\ref{TCvSHXEPS} and Fig.~\ref{TCvSHZEPS}, respectively. These results are able to be compared with Fig.~\ref{TcVHxEFTEPS} and Fig.~\ref{TcVHzEFTEPS}, that comes from EFT-1.
%
Looking to the Fig.~\ref{TCvSHXEPS}, we found that the critical temperature $T_c$ of the sublattice magnetization $m^{z}_A$ is identically to the \textit{melting temperature}. So, above $T_c$ the system is paramagnetic without any sublattice magnetization. The EFT-1 results shown in the Fig.~\ref{TcVHxEFTEPS} are qualitative equal to results of the LSWT-1, therefore both methods are in excellent agreement.

%

On the other hand, analyzing the figures Fig.~\ref{TCvSHZEPS} and Fig.~\ref{TcVHzEFTEPS}, which shows the behavior of the $T_c$ as a function of the $h_z$ for a fixed $h_x$, we can see a substantial difference between these two methods. Notice that, while the EFT-1 method found a monotonic behaviour for magnetization, the LSWT-1,2 suggests a more intriguing situation. The Fig.~\ref{TCvSHZEPS} indicates that the critical temperature of the two sublattice magnetizations possesses different behaviors, which depends on whether $h_z$ is less or greater than critical value $h_{xc}$, where the sublattice magnetizations go to zero. Below the critical point, the sublattice magnetization increases again and this occur because the polarized phase is always induced by the increasing of the $h_{z}$.

\section{Conclusion}
\label{conclusion}

In this paper, we investigated the quantum phase transitions and the existence of Bosonic Dirac points in the Ising model on a honeycomb lattice. The quantum phase transitions are induced by external fields.

Using two methods, LSWT and EFT-1, we determined the phase diagram in plane $(h_x,h_z)$ at zero temperature. From LSWT, we determine the bosonic quasiparticle excitation.
Depending of the magnetic phase, degenerated points can emerge within the spectrum at the $K$ points in the Brillouin zone. These points always occur along the critical line or within the polarized magnetic region. We demonstrated that around these special  points the spectrum is linear and is robust against a weak impurity and disorder. The dynamical critical exponent was also investigated leading to $z=1$, that indicates a linear crossing around these points.
These results often characterize Fermionic Dirac Materials and therefore can be used to classify our model as a Bosonic Dirac Material.

Following this, it is fascinate how the present simple model (inside the polarized phase) became a very strong candidate to describe a Bosonic Dirac material. As we see, this phase can be reached by a fine tune of the longitudinal and transverse fields. Which is very interesting, because in principle it makes possible to turn a non Bosonic Dirac material into a Bosonic Dirac material only adjusting the external parameters like magnetic fields. Beside that, the Bosonic Dirac materials should be robust against the magnon-magnon interaction as pointed by ref.~\cite{Balatsky}, from this perspective the existence of this Bosonic Dirac points depends only on the symmetry of the honeycomb lattice. Furthermore, differently from the graphene fermionic case, our bosonic quasiparticle excitation can also exhibit one additional linear behavior at the center of the BZ, a soft mode.

In the real space we found edge states for a lattice with zigzag termination, these edge states are a peculiar characteristic of the honeycomb lattice and can be observed in the AF and polarized phase. They proved to be robust against potential disorder and provides one more evidence for a Bosonic Dirac Material description.

To determine the temperature dependence and the possible quantum corrections, we calculating the magnetic properties using the LSWT-2 and EFT-1, which allow us to conclude that the methods are in great agreement.

It is worth to highlight yet the importance of the Bosonic Dirac points due its connection with production of spin wave excitations~\cite{Pires}, which can be used to breed efficient spin currents.  Accordingly, these excitations could find application inside of Spintronics and optical lattices~\cite{Offidani,Baltz,Gregersen}.

The real compound $Ba_2 Co Te O_6$ can be described by an effective Ising model on a honeycomb lattice, and accordingly the present study,
is a strong candidate to be a Bonsonic Dirac material \cite{Chanlert}.
Despite of the fact that this compound presents an approachable next-nearest-neighbor interaction $J_2$, the symmetry of honeycomb is preserved.
Another promise compound was syntetized by \cite{Okabe}.
It is a single crystal of the verdazy radical which also presents a Neel order that can be described by a spin 1/2 antiferromagntic Heisenberg model in a honeycomb lattice.

We expect to motivate experimental efforts to study the nature of the \emph{Magnon Bosonic Dirac points} and its relation with the quantum phase transitions in this simple model.

\section{Acknowledgment}

We would like to thank you the prof. Mucio Continentino and prof. Octavio D. Rodriguez Salmon for all assistance as well as the Brazilian agencies CNPq, Capes and Fapeam for providing all supporting.

\appendix

\section{The linear spin wave approach} \label{spinwave}

The LSWT often presents some advantages over the usual mean field approximations, once the classical ground state (GS) and the origin of fluctuations are identified, this treatment provides the calculation of the dispersion relation that can be connected with neutron spectroscopy measurements and so on taking to some experimental background~\cite{Pearson}.

For the present Hamiltonian Eq.~\ref{model}, in the absence of transverse and longitudinal fields, the classical ground state is  antiferromagnetic. However, when the fields are turned on, the transverse field is responsible for the quantum fluctuation~\cite{Dutta}. Previous works has been applied LSWT in the presence of a transverse fields~\cite{ColletaMila,Mila}. In this case, for low longitudinal fields, the transverse field deviates the spin into the $x$ direction. For this reason, the ground state, which should be considered is the canted Antiferromagnetic state (CAF). As result, these deviations produce a local rotation of the spin configurations over the $y$ axis, described by $\tilde{S}^{p}_{j}=S_{p} \mathcal{R}_{y}(S_{p}\phi_{p}) S^{p}_{j}$, where $p$ accounts the sublattice. Namely, the sublattices A and B stands for spin up and down. The ground state configuration is obtained by the minimum value for energy in respect to the angles $\phi_{A}$ and $\phi_{B}$. In order to characterize the expected classical ground states as a function of the spin angle configurations, we defined the canted CAF phase when $\phi_{A}\neq\phi_{B}$, and polarized when $\phi_{A}+\phi_{B}=\pi$.

After performing the rotation above on the Hamiltonian Eq.~\ref{model}, we can introduce the linear spin wave approach theory following Hostein-Primakoff transformation
\begin{eqnarray}\label{LSWT}
  \tilde{S}_{i}^{zp} &=& S - a^{\dag}_{i,p}a_{i,p} \\ \nonumber
  \tilde{S}_{i}^{xp} &=& \frac{\sqrt{2S}}{2}(a_{i,p} +a^{\dag}_{i,p}),
\end{eqnarray}
where, the $a^{\dagger}_{i,p}$($a_{i,p}$) is a bosonic creator (annihilator) operator of the vacuum state $| v \rangle$ and $S$ is the magnitude of the spin~\cite{Mila}. Substituting the spin wave mapping, Eq.~\ref{LSWT}, in the rotated Hamiltonian and performing a Fourier transformation, the quadratic Hamiltonian in a momentum space is finally obtained as

\begin{eqnarray}\label{modelSpinwave}
 \mathcal{H}^{LSWT} &=& E_c+\sum_{k}\tilde{A} a^{\dagger}_{k,A}a_{k,A} +\tilde{B} a^{\dagger}_{k,B}a_{k,B} \\ \nonumber
  &+& \sum_{k}(C_{-k}a_{k,A}a_{-k,B} + C_{-k}a_{k,A}a^{\dagger}_{k,B} + h.c.).
\end{eqnarray}
The energy of the classical state $E_c$ and the coefficients are given by
\begin{eqnarray}\label{ClassicalEnergy}
E_c&=&-\frac{NS}{2}[JzS\cos(\phi_A)\cos(\phi_B)+h_{x}(\sin(\phi_A)+\sin(\phi_B)) \nonumber \\
   &+&h_{z}(\cos(\phi_A)-\cos(\phi_B))]
\end{eqnarray}
\begin{equation}\label{coeffs}
\begin{cases}
& \tilde{A} = JzS\cos(\phi_{A})\cos(\phi_{B})+(h_x\sin(\phi_A)+h_z\cos(\phi_A)) \\
&\tilde{B} = JzS\cos(\phi_{A})\cos(\phi_{B})+(h_x\sin(\phi_B)-h_z\cos(\phi_B)) \\
& C_{k}=Jz\frac{S}{2}\sin(\phi_A)\sin(\phi_B)\gamma(k)
\end{cases},
\end{equation}
with $\gamma(k)=\frac{1}{z}\sum_{\delta}e^{k\delta}$ being the structure factor, $\delta$ a vector that localizes all first neighbors from a site i and $z$ is the number of neighbors.

So, in order to find the quasi-particles excitations and spectrum dispersion of the Magnons, we used the Bogoliubov-Valatin transformation~\cite{Mila} to obtain the following diagonal form of the Hamiltonian
\begin{equation}\label{modelSpinwaveDiagonal}
H^{LSWT} = E_g + \frac{1}{2}\sum_{k} (\omega_{1}(k) \alpha^{\dagger}_{k}\alpha_{k}+\omega_{2}(k) \beta^{\dagger}_{k}\beta_{k}  )
\end{equation}
such that,
\begin{equation}\label{Ground}
  E_g(T=0)=E_c+E_q
\end{equation}
is the ground state energy with the classical part $E_c$ and quantum correction equal to $E_q= -\frac{N}{2}(\tilde{A}+\tilde{B}) + \frac{1}{2}\sum_{k}(\omega_{1,k}+\omega_{2,k})$.

The quasiparticles energies is given by
\begin{eqnarray}\label{dispersion}
  \omega_{1,2}(k)&=&\frac{1}{2}\,\sqrt {2\,( {{\it \tilde{A}}}^{2}+{{\it \tilde{B}}}^{2} ) \pm 2 \,\sqrt{\Psi_k}},
\end{eqnarray}
where $\Psi_k=({{{\tilde{A}}}^{2}-{{\tilde{B}}}^{2}})^{2}+16{ \tilde{A}}{ \tilde{B}}{ |C_k|^{2}}$.

In special, for the honeycomb lattice $z=3$ and $\gamma(k)=\frac{1}{3}(\,{{\rm e}^{\frac{i}{2} \left( {\it kx a}+\,\sqrt {3}{\it ky a} \right) }}+\,{{\rm e}^{\frac{i}{2} \left( {\it kx a}-\,\sqrt {3}{\it ky a} \right) }}+\,{{\rm e}^{-i{\it kx a}}})$, with all summation and integrals defined over the first Brillouin zone. From here on, the lattice constant assumes $a=1$.

The sublattice magnetization along the rotated quantized axis can be calculated by the following expression
\begin{eqnarray}\label{magnetization}
 \bar{m}_p^z&=& S - \frac{1}{N_{uc}}\sum_{k}\left\langle a^{\dagger}_{k,p}a_{k,p}\right\rangle ,
\end{eqnarray}
with $N_ {uc} $ being the number of unity cells. In order to build the order parameter for each phase along the phase diagram in plane $h_x$-$h_z$, is convenient to define the magnetization along the old axis, i.e. $m^x_{p}=\bar{m}_p^z\sin(\phi_p)$ and $m^z_{p}=\bar{m}_p^z\cos(\phi_p)$.

\section{The low-energy description}\label{Lowenergy}

An effective low-energy description of the Hamiltonian can be obtained by a series expansion of $\gamma(k)$ around the critical points $\mathbf{k}_0$. These points are located on the center $\mathbf{\Gamma}$ and corners $\mathbf{K}$ of the Brillouin zone.
At the critical point $\phi_B =\pi-\phi_A$, that implies $\tilde{A}=\tilde{B}$ in Eq.~\ref{dispersion}. This constraint does not depend of BZ point choice.

Near of the $\mathbf{\Gamma}$ point, $\gamma(\mathbf{\Gamma}) \sim ( 1- \frac{k^2}{4} ) $, for $k^2=k_x^2+k_y^2$. Substituting this expression in Eq.~\ref{dispersion} (with $(+)$ within the first square), we get
\begin{equation}\label{Gama}
\omega(\mathbf{\Gamma}) \sim  D \sqrt{ \Delta^{2 \nu z } +  k^{2z} },
\end{equation}
the coefficient is given by $D=\sqrt{ \frac{{\tilde{A} J z S {\sin(\phi_A)}^2}}{4}}$. While the gap function can be written as $\Delta=\frac{\sin(\phi_A)^2}{D^2} \left( h_{x} - h_{xc} \right)$, since we choose to fix $h_{z}$ and looking for the critical value $h_{xc}=\frac{JzS-h_{z}\cos(\phi_{A})}{\sin(\phi_{A})}$ where the gap vanishes. The Eq.~\ref{Gama} was written as a function of the critical exponents, leading to $\nu=\frac{1}{2}$ and $z=1$ for the critical exponents of spatial and temporal correlation lengths. Note, $\omega(\mathbf{\Gamma})$ is linear ``if and only if" $\Delta =0$, which is true for instance, at the critical point $h_x = \frac{3}{2}J$, $h_z=0$ and $\phi_A=\frac{\pi}{2}$, that leads $h_{xc}=JzS=\frac{3}{2}J$ and the gap function goes to zero. Is useful to realize that the dynamical critical exponent $z=1$ indicates a linear behaviour of the energy dispersion at the critical point and near of the soft point $\mathbf{\Gamma}$. 

Now, performing the same but around of the $\mathbf{K}$ points, the series expansion for $\gamma(k)$ provides $ \gamma(\mathbf{K}+\delta k) \sim ( - \frac{\sqrt{3}}{4}\delta k_x - \frac{1}{4}\delta k_y ) + i ( \frac{1}{4}\delta k_x -  \frac{\sqrt{3}}{4}\delta k_y ) $, where $\delta k_x$ and $\delta k_y$ are low increments of $k$ in the vicinity of $K$. Substituting $\gamma(k)$ in Eq.~\ref{dispersion}, yields
\begin{equation}\label{Kpoint}
 \omega(\mathbf{K}) \sim \tilde{A} \pm  \frac{J z S \sin(\phi_A)^2}{4}  \delta k.
\end{equation}
In this case, the gap goes to zero at finite value of the energy dispersion given by $\tilde{A}$. Therefore the gap function can be written as
$ \Delta = 2 (\omega(\mathbf{K})-A_c) $, which renders $\nu=1$, $z=1$ and as a consequence an linear behaviour conclusion for $\omega(\mathbf{K})$. This result is one of the characteristics of a Bosonic Dirac Materials.

\pagebreak
\section{Hamiltonian in an real space}\label{HRealSpace}

In the momentum space $k=\{k_{x},k_{y}\}$, we have
\begin{equation}\label{Hkxky}
\mathcal{H}=E_{q}+\frac{1}{2}\sum_{k=\{k_{x},k_{y}\}}\psi^{\dag}\mathcal{\bar{H}}(k_{x},k_{y})\psi ,
\end{equation}
for a basis $\psi^{\dag}=\{a^{\dag}_{k}, b^{\dag}_{k}, a_{-k}, b_{-k}\}$ and
\begin{equation}\label{Hbarkxky}
\mathcal{\bar{H}}(k_{x},k_{y})=
\left(
  \begin{array}{cccc}
   \tilde{A}  &  C_{k} & 0 & C_{k} \\
   C_{-k}  & \tilde{B} & C_{-k} & 0 \\
   0  &  C_{k} & \tilde{A} & C_{k} \\
   C_{-k} & 0 & C_{-k} & \tilde{B} \\
  \end{array}
\right).
\end{equation}

After a Fourier transform to real space in $x$-axis, we need to define the $m$ index that runs over the $L_{x}$ pair of sublattice $A$ and $B$ along the now finite axis. Following this, each matrix element in \ref{Hbarkxky} becomes a matrix of $L_{x}$-order, that can be written as
\begin{equation}\label{FTkx}
\mathcal{\bar{H}}_{I,J}(k_{y})=\frac{1}{N_{x}}\sum_{k_{x}}e^{ik_{x}(m-m^{\prime})}\mathcal{\bar{H}}_{I,J}(k_{x},k_{y}),
\end{equation}
where, $I$ and $J$ identify a matrix element from \ref{Hbarkxky}.
Therefore, to complete bulk matrix in $x$-axis real space, $\mathcal{\bar{H}}(k_{y})$ must be of $4L_{x}$ order and the basis is increased as
\begin{eqnarray}\label{basis}
   \psi^{\dag}=&&\{ a^{\dag}_{1,k_{y}}\cdots a^{\dag}_{m,k_{y}},b^{\dag}_{1,k_{y}}\cdots b^{\dag}_{m,k_{y}}, \nonumber \\
    &&a_{1,-k_{y}}\cdots a_{m,-k_{y}}, b_{1,-k_{y}}\cdots b_{m,-k_{y}}\}.
\end{eqnarray}
Accordingly \ref{FTkx}, each matrix element in \ref{Hbarkxky} becomes a matrix as
\begin{eqnarray}\label{MMorder}
\tilde{A} && \longrightarrow \tilde{A}\delta_{m,m^{\prime}} \\ \nonumber
\tilde{B} && \longrightarrow \tilde{B}\delta_{m,m^{\prime}} \\ \nonumber
C_{k} && \longrightarrow \tilde{C}\delta_{m,m^{\prime}}+ D_{k_{y}}\delta_{m,m^{\prime}+1} \\ \nonumber
C_{-k} && \longrightarrow \tilde{C}\delta_{m,m^{\prime}}+ D_{k_{y}}\delta_{m,m^{\prime}-1},
\end{eqnarray}
for, $\tilde{C}=J\frac{S}{2}\sin(\phi_{A})\sin(\phi_{B})$ and $D_{k_{y}}=2\tilde{C}\cos(\frac{\sqrt{3}}{2}k_{y})$. 

In order to reproduce the Zig-Zag edge configuration of the honeycomb lattice, we rewrite the $\mathcal{\bar{H}}(k_{y})$ in an appropriate basis~\ref{ZZbasis}, starting (left-edge) with a sublattice B and finishing (right-edge) with the sublattice A.
%
%
The pair of operators $\{b^{\dag}_{0,k_{y}},b_{0,-k_{y}}\}$ and $\{a^{\dag}_{m,k_{y}},a_{m+1,-k_{y}}\}$, represent the left-edge and right-edge, respectively.
The matrix $\mathcal{\bar{H}}(k_{y})$ on the Zig-Zag basis must assume the matrix form of \ref{HbarkyZZ} with $2(2L_{x}+2)$ order
\begin{widetext}
\begin{equation}\label{ZZbasis}
   \psi^{\dag}_{zz}=\{ b^{\dag}_{0,k_{y}},a^{\dag}_{1,k_{y}},b^{\dag}_{1,k_{y}}\cdots a^{\dag}_{m,k_{y}},b^{\dag}_{m,k_{y}},a^{\dag}_{m+1,k_{y}},
   b_{0,-k_{y}},a_{1,-k_{y}},b_{1,-k_{y}}\cdots a_{m,-k_{y}},b_{m,-k_{y}},a_{m+1,-k_{y}}\},
\end{equation}
\begin{eqnarray}\label{HbarkyZZ}
\mathcal{\bar{H}}(k_{y})_{zz}&=&
\left\{ \tilde{A}\delta_{2(m+1),2(m+1)}+ \tilde{B}\delta_{2(m+1),2m+1}+\left[\tilde{C}\delta_{2(m+1),2(m+1)+1}+D_{k_{y}}\delta_{2m+1,2(m+1)}+h.c. \right] \right\}
_{(2L_{x}+2)}\otimes \sigma_{0} \nonumber \\
&&+ \left[\tilde{C}\delta_{2(m+1),2(m+1)+1}+D_{k_{y}}\delta_{2m+1,2(m+1)}+h.c. \right]_{(2L_{x}+2)}\otimes \sigma_{1}.
\end{eqnarray}
\end{widetext}
where, $\sigma_{0}$ and $\sigma_{1}$ are the Pauli matrix. Now, the $m$-index must runs from zero to $L_{x}$, to take into account the pair formed by the left and right edges.

Since \ref{HbarkyZZ} represents a bosonic system, to calculate their eigenvalues the diagonalization process should be performs over the dynamic matrix $M_{D}=I_{-}\mathcal{\bar{H}}(k_{y})_{zz}$, where
\begin{equation}\label{Im}
I_{-}=\left(
 \begin{array}{cc}
   I_{2L_{x}+2}  & 0_{2L_{x}+2}   \\
   0_{2L_{x}+2}  & -I_{2L_{x}+2}
 \end{array}
\right),
\end{equation}
and $I_{{2L_{x}+2}}$ is an identity matrix of $(2L_{x}+2)$-order, same for the zero matrix $0_{2L_{x}+2}$.

\section{Effective-field theory}\label{EfectiveFieldTheory}

In order develop nonzero temperature phase diagrams, for comparison with the LSWT, we use the Effective Field Theory 1 (EFT-1) approach.  This treatment consider clusters with a single central spin. The EFT-1 Hamiltonian for the present honeycomb lattice with coordination number $z$, is given by
\begin{equation}
\mathcal{H}_{1p}=\left( J\overset{z}{\sum\limits_{\delta }}S_{(1+\delta)p^{\prime}}^{z}-h_{z}\right)S_{1p}^{z}-h_{x}S_{1p}^{x},
\label{2}
\end{equation}
where, $p$ denotes the sublattices A and B. While, $p^{\prime}$ stands for an opposite sublattice respect to $p$.

From the Eq.~\ref{2} and using the approximate Callen-Suzuki relation~\cite{barreto}, we obtain the magnetizations $m_{p}=\langle S_{1p}^{z}\rangle$
\begin{equation}
m_{p}=\left\langle \frac{h_{z}-a_{1p}}{\sqrt{(h_{z}-a_{1p})^{2}+h_{x}^{2}}}\tanh
\beta \sqrt{(h_{z}-a_{1p})^{2}+h_{x}^{2}}\right\rangle,
\label{4}
\end{equation}
for $a_{1p}=J\overset{z}{\sum\limits_{\delta}}S_{(1+\delta )p^{\prime}}^{z}$.

Applying the identity $\exp (\alpha D_{x})F(x)=F(x+a)$ (with $D_{x}=\frac{\partial }{\partial x}$ is the differential operator) and the Van der Waerden
identity for the spin up and down (i.e., $\exp (aS_{i}^{z})=\cosh (a)+S_{i}^{z}\sinh (a)$), the Eq.~\ref{4} leads
\begin{equation}
m_{p}=\left\langle \prod_{\delta\neq 0}^{z}(\alpha
_{x}+S_{(1+\delta )p^{\prime}}^{z}\beta _{x})\right\rangle \left.
F(x)\right\vert _{x=0},
\label{6}
\end{equation}
with
\begin{equation}
F(x)=\frac{h_{z}-x}{\sqrt{(h_{z}-x)^{2}+h_{x}^{2}}}\tanh \beta\sqrt{(h_{z}-x)^{2}+h_{x}^{2}},
\label{8}
\end{equation}
for $\alpha _{x}=\cosh (JD_{x})$ and $\beta _{x}=\sinh (JD_{x})$. The Eq.~\ref{6} is expressed in terms of multiple spin correlation functions.
It is practically impossible to obtain all the spin correlation functions. For this reason,
a decoupling of the right-hand side in Eq.~\ref{6}) is needed, namely
\begin{equation}
\left\langle S_{iA}^{z} S_{jB}^{z} \ldots S_{lA}^{z}\right\rangle \backsimeq m_{A} m_{B}\ldots m_{A},
\label{8a}
\end{equation}
where $i\neq j \neq \ldots \neq l$ and $m_{p}=\left\langle S_{ip}^{z} \right\rangle $. Is worth to note, the approximation of Eq.~\ref{8a} neglects correlations between different spins. However, relations as $\left\langle \left( S_{ip}^{z} \right)^{2} \right\rangle =1$ are exact, while in the usual
MFA all the self- and multispin correlations are neglected. After this, the Eq.~\ref{6} can be written as
\begin{equation}
m_{p}=\sum_{q=0}^{z} A_{q}(T_{N},h_{z},h_{x})m_{p^{\prime}}^{q},
\label{9}
\end{equation}
with
\begin{equation}
A_{q}(T_{N},H_{z},h_{x})=\frac{z!}{q!(z-q)!}\alpha _{x}^{z-q}\beta _{x}^{q}\left.
F(x)\right\vert _{x=0}.
\label{11}
\end{equation}
The coefficients $A_{q}(T_{N},h_{z},h_{x})$ are obtained by using the relation $\exp (\alpha D_{x})F(x)=F(x+a)$.

We can define the uniform and staggered magnetizations as $m=\frac{1}{2}(m_{A}+m_{B})$ and $m_{s}=\frac{1}{2}(m_{A}-m_{B})$, respectively. Accordingly, near the critical point we have $m_{s}\rightarrow 0$ and $m\rightarrow m_{0}$. So, the sublattice magnetization $m_{A}$ expanded up to linear order in $m_{s}$ (order parameter) is given by
\begin{equation}
m_{A}=X_{0}(T_{N},h_{z},h_{x},m_{0})+X_{1}(T_{N},h_{z},h_{x},m_{0})m_{s},
\label{12}
\end{equation}
with
\begin{equation}
X_{0}(T_{N},h_{z},h_{x},m_{0})=\sum_{q=0}^{z} A_{q}(T_{N},h_{z},h_{x})m_{0}^{q},
\label{13}
\end{equation}
and
\begin{equation}
X_{1}(T_{N},h_{z},h_{x},m_{0})=-\sum_{q=0}^{z} qA_{q}(T_{N},h_{z},h_{x})m_{0}^{q-1}.
\label{14}
\end{equation}

The present technique works only for second-order transitions. Therefore, to study the phase diagram we just analyze the
Eq.~\ref{13}) in the limit of $m_{\mathbf{s}}\rightarrow 0$. Thus, once the second-order line is located and using the fact that
$m_{A}=m_{0}+m_{s}$ in Eq.~\ref{12}), at the critical point we get
\begin{equation}
X_{0}(T_{N},\widetilde{h}_{z},\widetilde{h}_{x},m_{0})=m_{0},
\label{B15}
\end{equation}
\begin{equation}
X_{1}(T_{N},\widetilde{h}_{z},\widetilde{h}_{x},m_{0})=1,
\label{B16}
\end{equation}
in which $m_{s}=0$, $\widetilde{h}_{x}=h_{x}/J$ and $\widetilde{h}_{z}=h_{z}/J$.

We can now obtain an analytical solution for the second-order transition, where $m_{s}$ is the order parameter, to describe the phase transition of the present model Eq.~\ref{model}.
The phases are identified following $m_{s}\neq 0$ (different sublattice magnetizations) for the antiferromagnetic phase and $m_{s}=0$ (equal sublattice magnetizations) for the polirized phase.

\end{document}